# Critical Behavior of $Sn_2P_2S_6$ and $Sn_2P_2(Se_{0.28}S_{0.72})_6$ Crystals under High Hydrostatic Pressures

Zapeka B., Kostyrko M., Martynyuk-Lototska I. and Vlokh R.

*Vlokh Institute of Physical Optics, 23 Dragomanov Street, 79005 Lviv, Ukraine, E-mail: vlokh@ifo.lviv.ua*

**Abstract**

Basing on the temperature dependences of optical birefringence for $Sn_2P_2S_6$ and $Sn_2P_2(Se_{0.28}S_{0.72})_6$ crystals subjected to hydrostatic pressures, we prove unambiguously that $Sn_2P_2S_6$ reveals a tricritical point on its $(p,T)$-phase diagram with the coordinates $(p,T)$ = (4.3 kbar, 259 K), so that the second-order phase transition transforms into the first-order one whenever the pressure increases above 4.3 kbar. We also find that increasing hydrostatic pressure applied to $Sn_2P_2(Se_{0.28}S_{0.72})_6$ leads to the change in the phase transition character from tricritical to first-order. Further increase in the pressure up to ~ 2.5 kbar imposes splitting of the first-order paraelectric-to-feroelectric phase transition into two phase transitions, a second-order paraelectric-to-incommensurate one and a first-order incommensurate-to-ferroelectric transition.

## 1. Introduction

$Sn_2P_2S_6$ and $Sn_2P_2(Se_{0.28}S_{0.72})_6$ crystals belong to a family of ferroelectric–semiconductor solid solutions with the general chemical formula $(Sn_{1-y}Pb_y)_2P_2(Se_xS_{1-x})_6$ [1–4], which are among the most interesting materials from the viewpoint of their critical behavior occurring in the course of structural phase transitions. These solid solutions are characterized by the appearance of tricritical and triple points on their concentration–temperature phase diagrams (abbreviated hereafter as $(x,T)$ or $(y,T)$) and pressure–temperature, or $(p,T)$-phase diagrams [4].

Up to recently, there has been a commonly accepted opinion that these crystals manifest a special Landau point, a so-called Lifshitz point, on their phase diagrams [1–3]. This should mean that a proper second-order paraelectric–ferroelectric phase transition with the change of point symmetry $2/m \leftrightarrow m$ splits and an intermediate incommensurate phase appears at the Lifshitz point where the wave vector of superstructural modulation approaches zero [3, 5–8]. The fact of existence of the Lifshitz point with the coordinates $(x,T)$ = (0.28, 284 K) on the $(x,T)$-diagram of $Sn_2P_2(Se_xS_{1-x})_6$ solid solution has long been deemed as well proven experimentally [2, 3]. However, our recent works [6, 9–12] have revealed no Lifshitz point (or line) on the $(p,T,x)$-phase diagram of the solid solutions men-tioned above. Instead, we have shown that a change in the order of the phase transition from second to first occurs under thermodynamic conditions that precede splitting of this phase transition. Thus, tricritical and triple points should rather exist in the phase space characterizing these solid solutions.

This picture has been proven satisfactory by the studies of pressure–temperature behavior of the acoustic wave velocity $v_{55}$ for the $Sn_2P_2S_6$ crystals. It manifests a deep minimum at the pressure 4.3 kbar which can only be explained by appearance of the tricritical point with the coordinates $(p,T) = (4.3$ kbar, 259 K) [10]. Except in the works [12, 13], the critical behavior of the mixed $Sn_2P_2(Se_xS_{1-x})_6$ crystals occurring with approaching this tricritical point on the $(p,T)$-phase diagram has not yet been studied, though the relevant critical exponents represent the most important characteristics of the structural phase transitions. In fact, basing on some older works [14–18] concerned with the critical exponent $\gamma$ of dielectric permittivity for the lead-containing solid solutions, one can conclude that the behavior of these materials on their $(p,T)$-phase diagram is close to tricritical. The pressure behavior of the critical exponent $\beta$ of the order parameter has been studied up to 2.3 kbar only, demonstrating that this exponent is not pressure-dependent in the region of 0–2.3 kbar [13, 19]. On the other hand, our work [9] has testified that, instead of the Lifshitz point, the tricritical point with the coordinates $(x,T) = (0.28, 284$ K) appears on the $(x,T)$-phase diagram of $Sn_2P_2(Se_{0.28}S_{0.72})_6$ at the normal pressure. The appearance of the tricritical point with the coordinates (0.28, 284 K) has been proven by a number of experimental results. These include the approach of the critical exponent $\beta$ to the value ¼, as well as the approach of the critical exponent $\alpha$ of heat capacity to ½ [4].

In the context of our discussion it would be appropriate to remind that the Lifshitz point is quite a rare phenomenon. Among solid state materials, it has been observed only in ferromagnetic crystals MnP [20, 21] where a so-called 'fan phase' appears between the ordered and normal phases. The authors of the theoretical work [22] have stated that the Lifshitz point is energetically disadvantageous for ferroelectric crystalline materials, and this seems to be true. Notice also that the fact of splitting of the ferroelectric phase transition occurring in $Sn_2P_2(Se_{0.28}S_{0.72})_6$ crystals at high pressures ($p \approx 2.3$ kbar) has been ascertained basing on the 'secondary' experimental results rather than 'direct' ones. These are observations of broadening of the anomaly of the acoustic wave velocity $v_{22}$ near the phase transition temperature point and the corresponding increase in the acoustic wave attenuation [6]. Nonetheless, these results have allowed us to suggest a consistent general $(p,T,x)$-phase diagram for the $(Sn_{1-y}Pb_y)_2P_2(Se_xS_{1-x})_6$ solid solutions [4, 6]. Still a lot of fine details of this phase diagram require their experimental confirmation. In particular, one expects the fact of splitting of the phase transition in $Sn_2P_2(Se_{0.28}S_{0.72})_6$ at $p \approx 2.3$ kbar to be confirmed. Moreover, it is desirable to prove the appearance of the tricritical point in $Sn_2P_2S_6$ crystals at $p = 4.3$ kbar, using direct studies of pressure dependence of the critical exponent of the order parameter. These are the main goals of the present work.

## 2. Experimental

We remind that the $Sn_2P_2S_6$ and $Sn_2P_2(Se_{0.28}S_{0.72})_6$ crystals are monoclinic. Under normal conditions they are characterized with the point symmetries m and 2/m, respectively. The crystallographic axis $b$ is perpendicular to the symmetry mirror plane, while the $a$ axis is almost parallel to the direction of spontaneous polarization vector in the ferroelectric phase.

In our optical studies, the light wave vector was parallel to the $c$ axis in the case of $Sn_2P_2(Se_{0.28}S_{0.72})_6$ and to the $a$ axis in the case of $Sn_2P_2S_6$ crystals. Notice that the optical indicatrix rotates around the $b$ axis only with changing temperature. In addition, since the paraelectric phase is centrosymmetric, the temperature behavior of birefringence below the Curie temperature is described by a quadratic Kerr effect, which implies that the birefringence in low-temperature phase should be independent of the sign of spontaneous polarization. This fact means that the birefringence is the same in the domains with the opposite polarizations. This allows one to use linearly polarized incident light in the experimental geometries mentioned above. We placed our samples between crossed polarizers in a so-called 'diagonal' position, when the principal axes of the Fresnel ellipsoid cross-section were rotated by $\pm 45$ deg with respect to the transmission directions of the polarizers.

Under the change by $\pi/2$ in the optical phase difference, the intensity of light emergent from the optical polarization system is changed from its minimum to a maximum (below we will refer to this method as a method of intensity oscillation). With the phase difference having changed by $\pi$, the emergent intensity changes from its minimal value – through a maximum – and then down to a next minimum, or vice versa [23, 24]. In such a case the increment of the optical birefringence $\delta(\Delta n)$ measured between two fixed temperatures that correspond to, e.g., minimal light intensities is equal to $\delta(\Delta n) = k\lambda/d$, where $d$ denotes the sample thickness, $k$ the number of minimums of the intensity, and $\lambda = 632.8$ nm the wavelength of a He-Ne laser used by us as a light source.

We prepared our samples in the shape of thin plates with the thicknesses 0.98 mm in case of $Sn_2P_2S_6$ and 0.48 mm in case of $Sn_2P_2(Se_{0.28}S_{0.72})_6$ crystals. They were polished to ensure optical quality of their surfaces. The birefringence increments were measured in the temperature range of $\sim 180$ K – 360 K in the heating or cooling runs. The $Sn_2P_2S_6$ sample was preliminary heated up to 360 K or cooled down to $\sim 215$ K. Then the birefringence increment was measured while the temperature was changing according to a natural cooling (or heating) mechanism. In case of $Sn_2P_2(Se_{0.28}S_{0.72})_6$ crystals, the sample was preliminary cooled down to $\sim 180$ K and then the birefringence increment was measured during a natural heating process. Although the temperature change rate inside a sample chamber was not constant under these circumstances, the average temperature scan rate was sufficiently low ($\partial T/\partial t \simeq 0.4$ K/min). The accuracy of the birefringence increment measurements was not worse than $5 \times 10^{-3}$.

The samples of $Sn_2P_2S_6$ and $Sn_2P_2(Se_{0.28}S_{0.72})_6$ were placed into a high-pressure optical chamber with optical windows made of fused quartz. The hydrostatic pressure was changed in the range of 0–4.5 kbar inside the chamber, using a pumping station UNGR-20000 and oil as a pressure transmitter.

According to our experimental procedures, the pressure was applied to the samples and the temperature-induced birefringence increments were measured at each fixed pressure.

## 3. Results and Analysis

Fig. 1 displays the temperature dependences of the birefringence increment measured for the $Sn_2P_2S_6$ and $Sn_2P_2(Se_{0.28}S_{0.72})_6$ crystals at different hydrostatic pressures. As seen from Fig. 1,a, the slopes of the temperature dependences corresponding to $Sn_2P_2S_6$ increase with increasing pressure. This gradual process is observed up to the pressure 4.5 kbar after which the slope change becomes more pronounced. The latter means that the critical exponent $\beta$ decreases with pressure increasing, beginning from the value 0.30±0.01 (see Refs. [4, 19]) at the atmospheric pressure. On the contrary, the pressure behavior of the birefringence increment for $Sn_2P_2(Se_{0.28}S_{0.72})_6$ crystals (see Fig. 1,b) is not so evident. It is only seen that the birefringence anomalies for these crystals are slightly diffused in the vicinity of the phase transition. It is obvious only that the Curie temperature for the both crystals decreases with pressure increasing. Moreover, a jump of the birefringence increment at $T_C$, which is peculiar for first-order phase transitions, cannot be excluded for $Sn_2P_2(Se_{0.28}S_{0.72})_6$. Indeed, it is difficult to determine the coefficient $k$ in the formula $\delta(\Delta n) = k\lambda/d$ whenever such a jump exists. Therefore, our further analysis requires more refined mathematical processing of the experimental data.

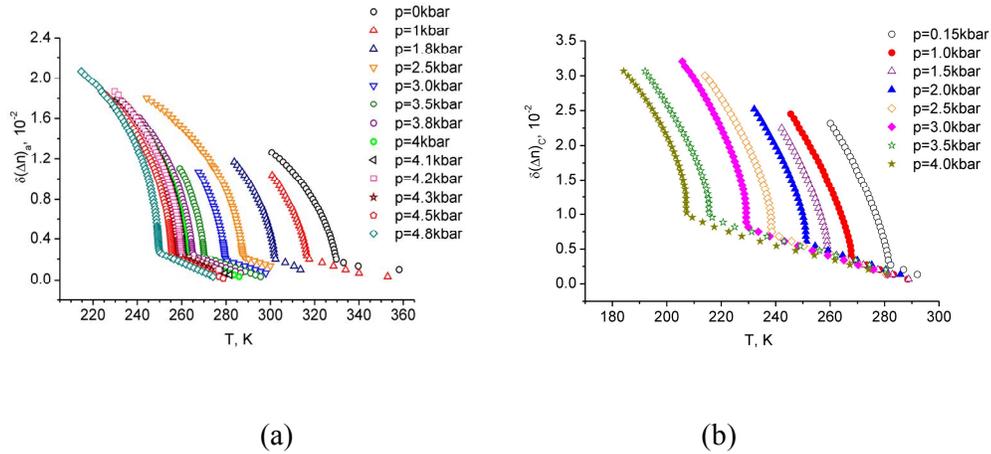

(a)          (b)

Figure 1. Temperature dependences of birefringence increment for (a) $Sn_2P_2S_6$ and (b) $Sn_2P_2(Se_{0.28}S_{0.72})_6$ crystals measured at different hydrostatic pressures.

The crystals under study exhibit rather high natural birefringence value and temperature rates of the birefringence, especially near the Curie temperature $T_C$. If the temperature scan rate were not low enough, the actual temperature (and so the birefringence) would have been inhomogeneously distributed in the sample. Notice that earlier we have found [19] that the acceptable temperature scan rates for which the anomalies near the $T_C$ point in $Sn_2P_2S_6$ crystals do not become diffused should be lower than ~ 0.12 K/min. Issuing from the diffused character of the birefringence anomalies near $T_C$, we cannot determine the exact $T_C$ values. As a consequence, the critical exponents can be found only in a rough approximation.

To solve the problem more accurately, we are to apply the approach developed in the work [25]. Namely, we consider a phase transition diffused due to some scalar inhomogeneity, e.g., scalar-type defects. The latter do not change the symmetry of a crystalline matrix, though their availability imposes some distribution of the Curie temperature $T_C$ over a sample (notice that this model is similar to that described in Ref. [26]). Then the phase transition temperature should be distributed over the sample within some temperature region $\Delta T = T_{CN} - T_{C1}$, where $T_{C1}$ and $T_{CN}$ are respectively the lowest and the highest Curie points. Now we divide the sample into $N$ homogeneous elementary cells (in our simulations we have taken $N$ to be equal to ten). The Curie temperature of the $i$ th cell is determined as $T_{Ci} = T_{CN} - i \times \Delta T / N$. Since the slope of temperature dependence of the birefringence also depends on this temperature inhomogeneity inside the sample, it is necessary to introduce the corrections to the critical exponents $B(T_{CN} - i\Delta T / N - T)$, where $B$ is some parameter. Let us introduce the birefringence change associated with the phase transition, which is determined as a difference of the actual birefringence in the ferroelectric phase and the birefringence value obtained as a linear interpolation from the paraelectric phase into ferroelectric one, i.e. as $\delta(\Delta n)_a^f - \delta(\Delta n)_a^p$. This parameter conventionally referred further on to as the 'birefringence difference' can be fitted using the formula (see, e.g., Ref. [19])

$$\delta(\Delta n)_a^f - \delta(\Delta n)_a^p = \frac{A}{N} \sum_{i=1}^{N} \left( T_{CN} - \frac{i\Delta T}{N} - T \right)^{2\beta + B(T_{CN} - \frac{i\Delta T}{N} - T)}. \tag{1}$$

Some of the temperature dependences of the birefringence differences fitted with formula (1) for the case of $Sn_2P_2S_6$ crystals are presented in Fig. 2 and the fitting parameters are gathered in Table 1.

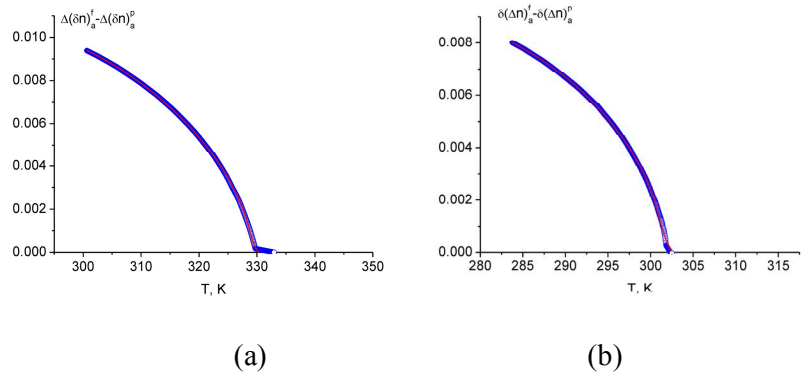

(a)                    (b)

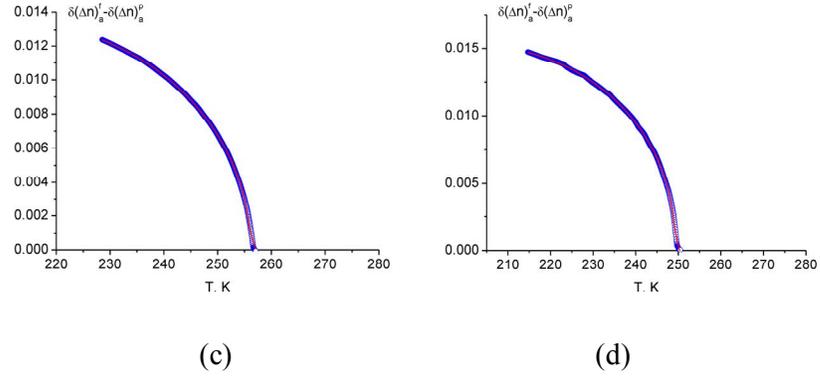

(c)                 (d)

Figure 2. Temperature dependences of birefringence differences for $Sn_2P_2S_6$ crystals obtained at different pressures: (a) 0.2, (b) 1.8, (c) 4.3, and (d) 4.8 kbar. Data points correspond to experiment and curves to fitting with formula (1).

Table 1. Fitting parameters for the temperature dependences of birefringence difference obtained for $Sn_2P_2S_6$ crystals.

| $p$, kbar | $A$, $10^{-3}$, K$^{-1}$ | $T_{CN}$, K | $\Delta T$, K | $\beta$ | $B$, $10^{-3}$, K$^{-1}$ | $R^2$ |
|---|---|---|---|---|---|---|
| 0.2 | 1.42±0.006 | 330.0±0.01 | 2.0 | 0.31±0.01 | 1.57±0.003 | 0.99989 |
| 1.0 | 1.52±0.03 | 317.4±0.03 | 2.0 | 0.31±0.03 | 1.99±0.0004 | 0.99663 |
| 1.8 | 1.86±0.007 | 302.8±0.01 | 2.0 | 0.28±0.03 | 2.29±0.06 | 0.99982 |
| 2.5 | 2.14±0.009 | 288.4±0.01 | 2.0 | 0.28±0.01 | 1.77±0.003 | 0.99986 |
| 3.0 | 2.11±0.007 | 280.4±0.01 | 2.0 | 0.28±0.03 | 1.59±0.12 | 0.99986 |
| 3.5 | 2.32±0.01 | 270.9±0.01 | 2.0 | 0.28±0.05 | 2.68±0.19 | 0.99976 |
| 3.8 | 2.41±0.006 | 265.7±0.01 | 2.0 | 0.28±0.02 | 2.28±0.003 | 0.99995 |
| 4.0 | 2.86±0.006 | 262.7±0.01 | 2.0 | 0.23±0.01 | 1.50±0.004 | 0.99991 |
| 4.1 | 2.87±0.007 | 261.8±0.01 | 2.0 | 0.25±0.01 | 1.70±0.002 | 0.99993 |
| 4.2 | 3.04±0.01 | 260.4±0.01 | 2.0 | 0.24±0.01 | 1.27±0.003 | 0.99985 |
| 4.3 | 2.82±0.008 | 257.3±0.01 | 2.0 | 0.25±0.01 | 1.62±0.002 | 0.99990 |
| 4.5 | 3.12±0.007 | 256.3±0.01 | 2.0 | 0.24±0.01 | 1.64±0.002 | 0.99994 |
| 4.8 | 3.56±0.008 | 250.5±0.01 | 2.0 | 0.23±0.01 | 1.47±0.002 | 0.99993 |

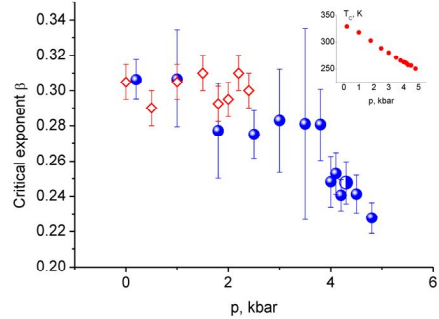

Figure 3. Dependence of critical exponent $\beta$ on the hydrostatic pressure for $Sn_2P_2S_6$ crystals: balls correspond to the data of the present work, diamonds to those of our recent work [19], and a semi-open circle to the pressure 4.3 kbar (with the critical exponent being 0.25). The insert shows dependence of phase transition temperature $T_C$ on the pressure.

The relevant data testify that the critical exponent decreases gradually with increasing pressure down to 0.25±0.01 (in the pressure range 4.1–4.5 kbar – see Fig. 3). This corresponds to thermodynamic conditions of the tricritical point. The result agrees satisfactorily with the data of our study of the temperature behavior of acoustic parameters under the hydrostatic pressure [10], where the tricritical point has been revealed at the pressure ~ 4.3 kbar. Moreover, there is a clear agreement with our results for the pressure dependence of the critical exponent found for the region of 0–2.3 kbar [19]. In fact, the critical exponent at 4.3 kbar is equal to 0.25 (see Fig. 3). This exponent acquires the magnitudes less than 0.25 under the pressures above ~ 4.5 kbar, which has no physical meaning from the standpoints of the classical Landau theory. Obviously, the pressures higher than the above value should correspond to thermodynamic conditions under which a first-order phase transition occurs. Finally, the dependence of the Curie temperature on the hydrostatic pressure for $Sn_2P_2S_6$ crystals is linear (see Fig. 3, insert).

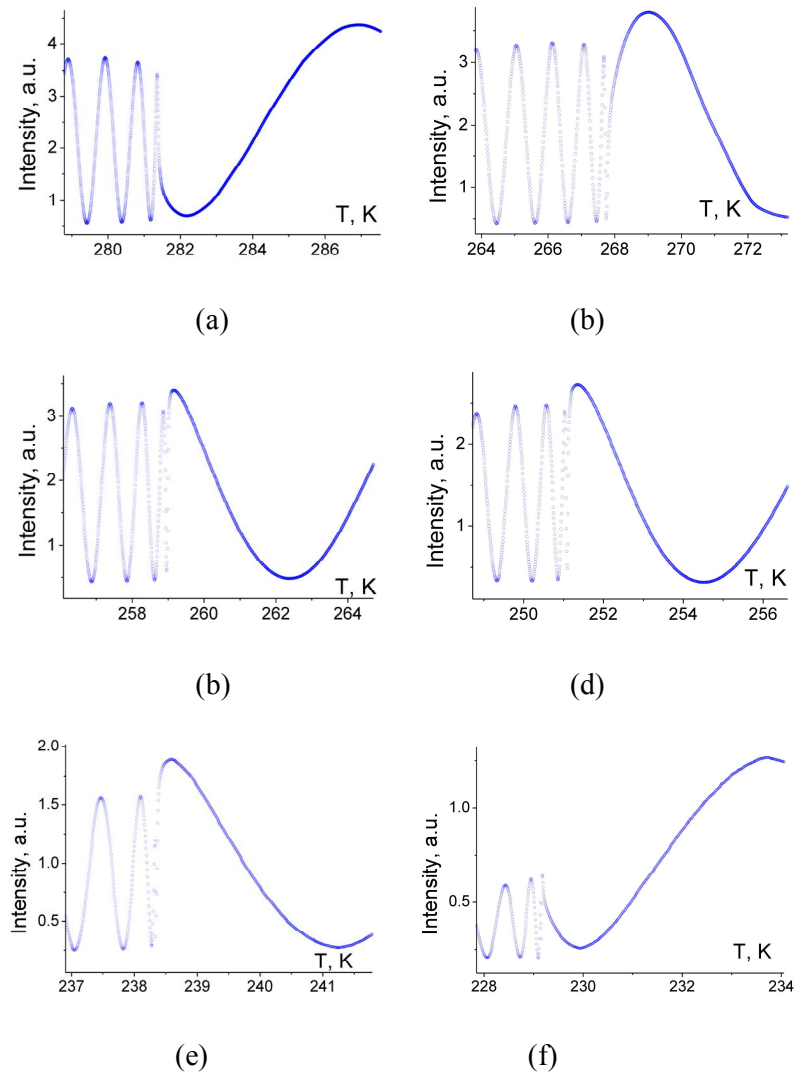

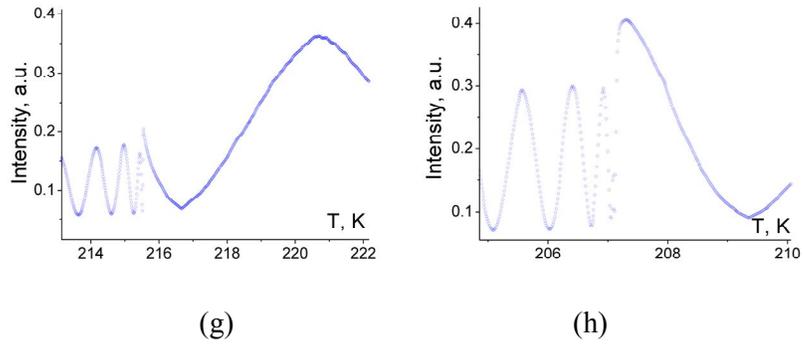

(g)                          (h)

Figure 4. Temperature dependences of intensity of light emergent from the system of crossed polarizers and $Sn_2P_2(Se_{0.28}S_{0.72})_6$ sample placed in between, as obtained under different hydrostatic pressures: (a) 0.15, (b) 1.0, (c) 1.5, (d) 2.0, (e) 2.5, (f) 3.0, (g) 3.5, and (h) 4.0 kbar.

In order to detect splitting of the ferroelectric phase transition in $Sn_2P_2(Se_{0.28}S_{0.72})_6$ compound happening under high pressures, we have studied the oscillations of light intensity emergent from our optical polarization scheme. Examples of the relevant temperature dependences for different hydrostatic pressures are presented in Fig. 4. Below the pressure point ~ 2.0 kbar, the oscillating temperature dependences of the light intensity reveal the only point where the frequency and the phase of the oscillations change abruptly (see Fig. 4,a–d). Of course, these temperature points correspond to the paraelectric–ferroelectric phase transition.

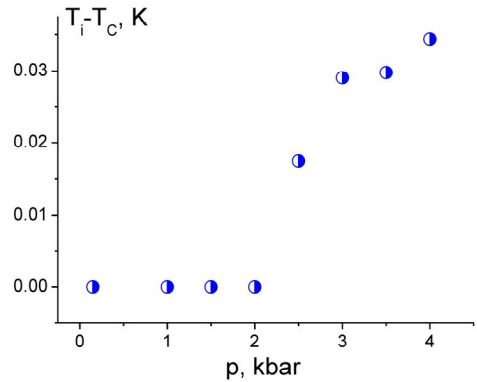

Figure 5. Pressure dependence of temperature width of the incommensurate phase in $Sn_2P_2(Se_{0.28}S_{0.72})_6$ crystals.

As seen from Fig. 4,e–h, two such temperature points appear in the pressure region 2.5–4.0 kbar. They can be termed as the temperatures of incommensurate–ferroelectric and paraelectric–incommensurate phase transitions, $T_C$ and $T_i$. This behavior of the oscillation parameters implies that there are three different temperature regions characterized by different temperature slopes of the birefringence. We are inclined to believe that these regions correspond to different structural phases, i.e. the normal phase at $T > T_i$, the intermediate phase at $T_C < T < T_i$ (most probably, incommensurate), and the ordered ferroelectric phase at $T < T_C$. Notice that, according to our results on the acoustic wave velocity and attenuation [6], splitting of the ferroelectric–paraelectric phase transition has to appear at ~ 2.3 kbar. This agrees perfectly with the present results. As seen from Fig. 5, the temperature width of the incommensurate phase is relatively narrow and increases with increasing pressure.

Table 2. Fitting parameters for the temperature dependences of birefringence difference obtained for $Sn_2P_2(Se_{0.28}S_{0.72})_6$ crystals.

| $p$, kbar | $A$, $10^{-3}$ K$^{-1}$ | $T_{iN}$, K | $\Delta T$, K | $\beta$ | $J$, $10^{-3}$ | $B$, $10^{-4}$ K$^{-1}$ | $R^2$ |
|---|---|---|---|---|---|---|---|
| 0.15 | 4.67±0.13 | 284.24±0.01 | 2 | 0.25±0.01 | −5.33±0.01 | 5.00±1.45 | 0.9998 |
| 1.0 | 4.74±0.13 | 270.51±0.01 | 2 | 0.240±0.004 | −4.86±0.02 | 5.00±0.85 | 0.9999 |
| 1.5 | 4.65±0.12 | 261.10±0.01 | 2 | 0.241±0.004 | −3.29±0.01 | 5.00±1.26 | 0.9997 |
| 2.0 | 5.56±0.09 | 253.33±0.01 | 2 | 0.215±0.003 | −4.02±0.01 | 5.00±0.59 | 0.9996 |
| 2.5 | 7.92±0.27 | 241.11±0.02 | 2 | 0.186±0.004 | −7.23±0.03 | 1.00±0.65 | 0.9996 |
| 3.0 | 9.80±0.45 | 231.90±0.02 | 2 | 0.17±0.01 | −8.54±0.05 | 3.07±0.79 | 0.9994 |
| 3.5 | 9.33±0.71 | 219.11±0.03 | 2 | 0.17±0.01 | −10.00±0.07 | 5.00±1.04 | 0.9992 |
| 4.0 | 9.54±0.37 | 209.46±0.01 | 2 | 0.15±0.01 | −7.02±0.05 | 2.73±0.74 | 0.9997 |

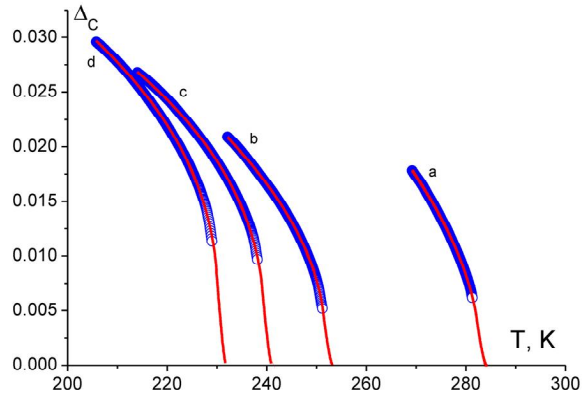

Figure 6. Temperature dependences of birefringence difference for $Sn_2P_2(Se_{0.28}S_{0.72})_6$ crystals obtained at different pressures: (a) 0.15, (b) 2.0, (c) 2.5, and (d) 3.0 kbar. Data points correspond to experiment and curves to fitting with formula (1).

Now we analyze the birefringence increment $\Delta_C = \delta(\Delta n)_C^f + \delta(\Delta n)_C^i - \delta(\Delta n)_C^p$ (with $\delta(\Delta n)_C^i$ being the birefringence increment in the incommensurate phase) calculated for the phase transition in the $Sn_2P_2(Se_{0.28}S_{0.72})_6$ crystals (see Fig. 6 and Table 2). To take into account a possible birefringence jump occurring at the first-order incommensurate–ferroelectric phase transition (or the paraelectric–ferroelectric phase transition), together with a temperature dependence of the birefringence in the incommensurate phase, we have modified relation (1) via including the additional term $J$, which is responsible for shifting up of the gradual birefringence dependence:

$$\Delta_C = \frac{A}{N}\sum_{i=1}^{N}\left(T_{iN} - \frac{i\Delta T}{N} - T\right)^{2\beta + B(T_{iN} - \frac{i\Delta T}{N} - T)} + J. \quad (2)$$

Instead of $T_{CN}$, here we denote the temperature of the commensurate–incommensurate phase transition as $T_{iN}$. The shift mentioned above can appear due to a birefringence jump (or very rapid change in the case of diffused anomalies) in the course of first-order phase transitions. Of course, this jump cannot be detected reliably while using the light intensity oscillation technique. From other side, as it has been found by us [27], at the atmospheric pressure the measured increment of birefringence $\Delta_C$ in $Sn_2P_2(Se_{0.28}S_{0.72})_6$ crystals, for example at the temperature 275 K, is equal to 0.018. At the same time due to the present results this value is smaller and is equal to 0.0081. It is seen from Table 2 that –$J$=0.0053 at the atmospheric pressure. Thus the total birefringence increment with accounting of value of jump is equal to 0.0134 and is much closer to the value obtained in work [27].

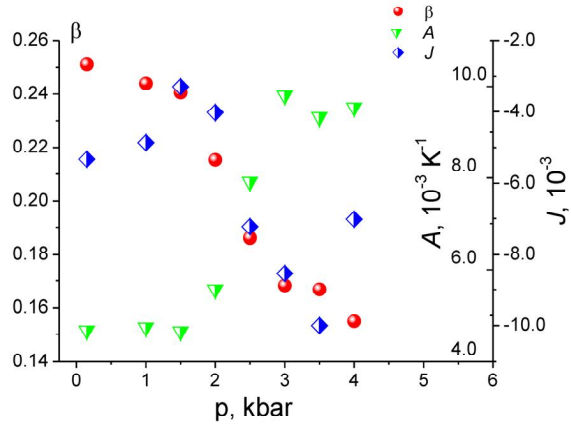

Figure 7. Pressure dependences of the critical exponent $\beta$ and the fitting parameters $A$ and $J$.

As seen from Fig. 7 and Table 2, the critical exponent of the order parameter for $Sn_2P_2(Se_{0.28}S_{0.72})_6$ decreases with increasing pressure and becomes less than 0.25, which is physically unrealistic from the viewpoint of the Landau theory. This may be caused by passing, in the phase space, into the region of first-order phase transitions, as well as by splitting of the paraelectric–ferroelectric phase transition. At the same time, the absolute value of the parameter $J$ which in fact represents the birefringence jump at the first-order phase transition increases rapidly at the pressures higher than ~ 2.5 kbar (see Fig. 7). Fig. 7 shows that the parameter $A$ also notably increases almost at the same pressure. In other words, the three parameters mentioned above manifest anomalous behaviors in the vicinity of the pressure point ~ 2.5 kbar. This fact indicates that the pressure ~ 2.5 kbar corresponds to a special thermodynamic coordinate at which the paraelectric–ferroelectric phase transition splits into the two transitions, the paraelectric–incommensurate and incommensurate-ferroelectric ones. Notice also that the latter conclusion agrees well with the results obtained in our recent study [6]. Of course, the final conclusion on the appearance of the incommensurately modulated phase under the high pressures in $Sn_2P_2(Se_{0.28}S_{0.72})_6$ crystals can be made only on the basis of structural data.

## 4. Conclusion

The studies of temperature variations of the optical birefringence of $Sn_2P_2S_6$ crystals under hydrostatic pressure have demonstrated that the critical exponent $\beta$ of the order parameter decreases from the value 0.31±0.01 at the atmospheric pressure down to the value 0.25±0.01 at 4.3 kbar, thus indicating that the tricritical point with the coordinates $(p,T) = (4.3$ kbar, 259 K) appears on the $(p,T)$-phase diagram of these crystals. This is why the order of the phase transition in $Sn_2P_2S_6$ should change from second to first with increasing pressure above 4.3 kbar.

We have also found that increase in the hydrostatic pressure applied to $Sn_2P_2(Se_{0.28}S_{0.72})_6$ crystals leads to decreasing critical exponent $\beta$ below 0.25. This fact points to changing character of the phase transition in these crystals from tricritical to first-order. Further increase in the pressure results in anomalies of the critical exponent $\beta$, the empirical coefficient $A$ and the birefringence jump $J$ observed at the pressure ~ 2.5 kbar. On the basis of temperature oscillations of the emergent light intensity we have found that a single spike seen in the oscillation phase dependence at lower pressures is split into two spikes at the pressures above ~ 2.5 kbar. In our opinion, all of the facts mentioned above suggest unambiguously that the pressure point ~ 2.5 kbar corresponds to the thermodynamic coordinate at which the first-order ferroelectric–paraelectric phase transition is split into two phase transitions, the second-order paraelectric–incommensurate one and the first-order incommensurate–ferroelectric transition.

## Acknowledgement

We thank Prof. Yu. Vysochanskii for useful discussions and supplying the crystals. The authors acknowledge also financial support of the present study by the Ministry of Education and Science of Ukraine under the Project # 0111U010236.

## References

1. V. Yu. Slyvka and Yu. M. Vysochanskii. Ferroelectrics of $Sn_2P_2S_6$ family. Properties around the Lifshitz point. Uzhgorod: Zakarpattya, 1994.
2. Y. M. Vysochanskii, T. Janssen, R. Currat, R. Folk, J. Banys, J. Grigas and V. Damulionis. Phase transitions in ferroelectric phosphorous chalcogenide crystals. Vilnius: Vilnius University Publishing House, 2006.
3. V. Yu. Vysochanskii and Yu. M. Slivka. The Lifshitz point on the state diagrams of ferroelectrics. Sov. Phys. Uspekhi. 162 (1992) p. 139.

4. R. O. Vlokh, O. G. Mys, I. Yu. Martynyuk-Lototska, B. Ye. Zapeka and M. Ye. Kostyrko. Phase diagrams of ferroelectric solid solutions $Sn_2P_2(Se_xS_{1-x})_6$. Lviv: Institute of Physical Optics, 2013.
5. L. Landau. On the theory of phase transitions. Ukr. J. Phys. (Special Issue) 53 (2008) p. 25.
6. O. Mys, I. Martynyuk-Lototska, B. Zapeka and R. Vlokh. On the special points on (P, T, x)-phase diagram for $Sn_2P_2(Se_xS_{1-x})_6$ crystals: acoustic studies of $Sn_2P_2(Se_{0.28}S_{0.72})_6$. Phil. Mag. 93 (2013) p. 648.
7. O. Andersson, O. Chobal, I. Rizak and V. Rizak. Tricritical Lifshitz point in the temperature–pressure–composition diagram for $(Pb_ySn_{1-y})_2P_2(Se_xS_{1-x})_6$ ferroelectrics. Phys. Rev. B. 80 (2009) 174107 (5 pp.).
8. A. Kohutych, R. Yevych, S. Perechinskii, V. Samulionis, J. Banys and Yu. Vysochanskii. Sound behavior near the Lifshitz point in proper ferroelectrics. Phys. Rev. B. 82 (2010) 054101 (10 pp.).
9. I. Martynyuk-Lototska, O. Mys, B. Zapeka and R. Vlokh. About the existence of a Lifshitz point on the phase diagram of $Sn_2P_2(Se_xS_{1-x})_6$ solid solutions: acoustic and optical studies. Phil. Mag. 91 (2011) p. 3519.
10. I. Martynyuk-Lototska, A. Say, O. Mys and R. Vlokh. On the tricritical point on (p, T)-phase diagram of $Sn_2P_2S_6$ crystals. Phil. Mag. 91 (2011) p. 4293.
11. I. Martynyuk-Lototska, A. Say, B. Zapeka, O. Mys and R. Vlokh. On the (P, T, x)-phase diagram of $Sn_2P_2(Se_xS_{1-x})_6$ solid solutions: comparison of temperature dependences of the acoustic wave velocities in $Sn_2P_2S_6$ and $Sn_2P_2(Se_{0.28}S_{0.72})_6$ crystals under high hydrostatic pressures. Ukr. J. Phys. Opt., Suppl. 3. 13 (2012) p. S1.
12. Yu. Vasylkiv, O. Mys, B. Zapeka and R. Vlokh. On the critical exponent of order parameter in $Sn_2P_2S_6$ and $Sn_2P_2(Se_{0.28}S_{0.72})_6$ crystals. Ukr. J. Phys. Opt, 13 (2012) p. 135.
13. A. Say, O. Mys, D. Adamenko, A. Grabar, Yu. Vysochanskii, A. Kityk and R. Vlokh. Critical exponents of phase transition in ferroelectric $Sn_2P_2S_6$: comparison of optical and dilatometric data. Phase Trans. 83 (2010) p. 123.
14. A. G. Slivka, E. I. Gerzanich, P. P. Guranich, V. S. Shusta and M. I. Gurzan. Critical behavior of spontaneous polarization in $Sn_2P_2(Se_xS_{1-x})_6$ crystals near the Lifshitz point induced by hydrostatic pressure. Izv. AN SSSR, Ser. Fiz. 51 (1987) p. 2162.
15. P. P. Guranich, E. I. Gerzanich, V. S. Shusta and A. G. Slivka. Phase (P,T,x)-diagram of the ferroelectric crystals $(Pb_xSn_{1-x})_2P_2Se_6$ with the incommensurate phase. Fiz. Tverd. Tela. 30 (1988) p. 1189.
16. A. G. Slivka, E. I. Gerzanich, Yu. I. Tyagur and M. I. Gurzan. Phase (P,T)-diagram of $Sn_2P_2Se_6$ ferroelectrics. Fiz. Tverd. Tela. 27 (1985) p. 526.
17. V. S. Shusta, E. I. Gerzanich, A. G. Slivka and P. P. Guranich. Dielectric properties and (P,T)-diagrams of ferroelectric solid solutions $(Pb_xSn_{1-x})_2P_2S_6$. Ukr. Fiz. Zhurn. **34** (1989) p. 1855.

18. V. S. Shusta, E. I. Gerzanich, A. G. Slivka and P. P. Guranich. Phase (P,T,x)-diagram of the ferroelectric crystals $(Pb_xSn_{1-x})_2P_2S_6$. Fiz. Tverd. Tela. **31** (1989) p. 308.
19. B. Zapeka, Yu. Vasylkiv, O. Mys and R. Vlokh. On the tricritical behaviour of the $Sn_2P_2S_6$ crystals, Phase Trans. 84 (2011) p. 193.
20. C. C. Beccera, Y. Shapira, N. F. Oliveira and T. S. Chang. Lifshitz point in MnP. Phys. Rev. Lett. 44 (1980) p. 1692.
21. Y. Shapira, C. C. Beccera, N. F. Oliveira and T. S. Chang. Phase diagram, susceptibility, and magnetostriction in MnP: Evidence for a Lifshitz point. Phys. Rev. B. 24 (1981) p. 2780.
22. T. A. Aslanyan and A. P. Levanyuk. Critical Lifshitz points. Fiz. Tverd. Tela. 20 (1978) p. 804.
23. M. Born and E. Wolf. Principles of optics: Electromagnetic theory of propagation, interference and diffraction of light. Oxford: Pergamon Press, 1964.
24. I.G. Wood, B. Welber, W.I.F. David and A.M. Glazer. Ferroelastic phase transition in $BiVO_4$. II. Birefringence at simultaneous high pressure and temperature. J. Appl. Cryst. 13 (1980) p. 224.
25. D. I. Adamenko, I. M. Klymiv, Y. Vasylkiv and R. O. Vlokh. Optical activity and critical exponent of the order parameter in lead germanate crystals. 1. The case of diffused phase transition in $Pb_5Ge_3O_{11}$ doped with Cu, Ba and Si ions. Ukr. J. Phys. Opt. 10 (2009) p. 182.
26. G. A. Smolenskiy. Physics of ferroelectric phenomena. Leningrad: Nauka, 1985.
27. B. Zapeka, O. Mys and R. Vlokh. On the order of phase transition in $Sn_2P_2(Se_{0.28}S_{0.72})_6$ solid solutions: optical birefringence studies. Ferroelectrics. 418 (2011) p. 143.